\documentclass[letterpaper]{mn2e}
\input{psfig}
\def\beq{\begin{equation}}
\def\eeq{\end{equation}}
\def\barray{\begin{eqnarray}}
\def\earray{\end{eqnarray}}

\def\gtsima{$\; \buildrel > \over \sim \;$}
\def\ltsima{$\; \buildrel < \over \sim \;$}
\def\prosima{$\; \buildrel \propto \over \sim \;$}
\def\gsim{\lower.7ex\hbox{\gtsima}}
\def\lsim{\lower.7ex\hbox{\ltsima}}
\def\simgt{\lower.7ex\hbox{\gtsima}}
\def\simlt{\lower.7ex\hbox{\ltsima}}
\def\simpr{\lower.7ex\hbox{\prosima}}
\def\la{\lsim}
\def\ga{\gsim}

\newcommand{\apj}{ApJ}
\newcommand{\apjs}{ApJS}
\newcommand{\apjl}{ApJL}
\newcommand{\aj}{AJ}
\newcommand{\mnras}{MNRAS}
\newcommand{\aap}{A\&A}

\newdimen\hssize
\newdimen\hdsize
\def\fsdb07{f_{\rm s,D}}

\def\lesssim{\mathrel{\hbox{\rlap{\hbox{\lower4pt\hbox{$\sim$}}}\hbox{$<$}}}}
\def\gtrsim{\mathrel{\hbox{\rlap{\hbox{\lower4pt\hbox{$\sim$}}}\hbox{$>$}}}}

\def\aj{AJ}
\def\apj{ApJ}
\def\apjl{ApJ}
\def\apjs{ApJS}
\def\aap{A\&A}
\def\mnras{MNRAS}
\usepackage{amsmath}
\usepackage{graphicx}  
\usepackage{color}

\newcommand{\comment}[1]{}

\begin{document}


\title[A fundamental problem in low mass galaxy evolution]{A fundamental
problem in our understanding of low mass galaxy evolution}
\author[S. M. Weinmann et al.]
       {\parbox[t]{\textwidth}{
        Simone
        M. Weinmann$^{1}$\thanks{E-mail:weinmann@strw.leidenuniv.nl}, 
         Anna Pasquali$^{2}$, Benjamin D. Oppenheimer$^{1}$,  \\
 Kristian Finlator$^{3,4}$, J. Trevor
        Mendel$^{5}$, Robert A. Crain$^{1}$, Andrea V. Macci\`{o}$^{6}$
        }\\
\vspace*{3pt}\\
$^1$Leiden Observatory, Leiden University, P.O. Box 9513, 2300 RA
Leiden, The Netherlands\\
$^2$Astronomisches Rechen-Institut, Zentrum f\"{u}r Astronomie der 
Universit\"{a}t Heidelberg, M\"{o}nchhofstr. 12-14, 69120 Heidelberg,
Germany\\
$^3$Department of Physics, University of California, Santa Barbara, CA
93106, USA\\
$^4$Hubble Fellow\\
$^5$Department of Physics and Astronomy, University of Victoria,
Victoria,
British Columbia, V8P 1A1, Canada\\
$^{6}$Max-Planck-Institut f\"{u}r Astronomie, K\"{o}nigstuhl 17, 69117 Heidelberg, Germany\\
}


\date{}

\pubyear{2012}

\maketitle

\label{firstpage}


\begin{abstract}
Recent studies have found a dramatic difference between the observed number
density evolution of low mass galaxies and that predicted by
semi-analytic models. Whilst models accurately reproduce
 the $z=0$ number
density, they require that the evolution occurs rapidly at early
times, which is incompatible with the strong late evolution
found
in observational results.
 We report here the same discrepancy in two state-of-the-art cosmological hydrodynamical 
simulations, which is evidence that the problem is fundamental.
We search for the underlying cause of this problem using two complementary methods. Firstly, we
consider a narrow range in stellar mass of log($M_{\rm
  star}/(h^{-2}M_{\odot}$))=9 - 9.5 and look for
evidence of a different history of today's low mass galaxies
in models and observations. We find that the exclusion
of satellite galaxies from the analysis brings the median ages
and star formation rates of galaxies into reasonable agreement. However, the
models yield too few young, strongly star-forming galaxies.
Secondly, we construct a toy model to link the observed evolution of
specific star formation rates with the evolution of
 the galaxy stellar mass function.
We infer from this model that a key problem
in both semi-analytic and hydrodynamical models
 is the presence of a 
positive instead of a negative correlation between specific star formation rate and stellar
mass. 
 A similar positive correlation is found between the specific dark matter halo
accretion rate and the halo mass, indicating that model
galaxies are growing in a way that follows the growth of their host
haloes
too closely. It therefore appears necessary
to find
a mechanism that decouples the growth of low mass galaxies, which occurs primarily at late times,
from the growth of their host haloes, which occurs primarily at early times. We argue that the current
form of star-formation driven feedback implemented in most galaxy
formation models is unlikely to achieve this goal, owing to its
fundamental dependence on host halo mass and time.

\end{abstract}




\section{Introduction}
\label{sec:intro}

In recent years, models of galaxy formation and evolution have made substantial progress in explaining
the observed properties  of massive galaxies in the Universe over cosmic epochs. 
This is due both
to the inclusion of AGN feedback in the models (e.g. Di Matteo et al. 2005, Bower et al. 2006, 
Croton et al. 2006; De Lucia et al. 2006), and a better understanding of the assembly history
of massive galaxies (e.g. Neistein et al. 2006).

Only very recently has it become clear that \emph{a fundamental problem 
with low mass galaxy evolution exists in these models}, at
$\log(M_{\rm star}/{\rm M_{\odot}}) \sim 8-10$, 
challenging the current models of galaxy evolution. This is the mass range
in which feedback by supernovae, stellar winds and stellar radiation
pressure, which remains  poorly understood, is believed to have a crucial impact on  galaxy evolution
(e.g. White \& Rees 1978; White \& Frenk 1991; Somerville \& Primack 1999; Benson et al. 2003). 
Problems with low mass galaxies have been identified, in various forms, in both semi-analytic models
and hydrodynamical simulations, as we discuss below.

Semi-analytic models
that include strong stellar feedback  accurately reproduce the $z=0$ 
stellar mass function (e.g. Guo et al. 2011; Bower, 
Benson \& Crain 2012), but they 
consistently build up this mass function too early, thus overproducing
the sub-M* mass function at $z>0.5$
(e.g. Fontana et al. 2006; Fontanot et al. 2007, 2009; Marchesini et al. 2009; Lo Faro et al. 2009; Guo et al. 2011).
In addition, there are indications that low mass galaxies at $z=0$ 
are too passive (e.g. Fontanot et al. 2009; Firmani \& Avila-Reese et al. 2010; Guo et al. 2011), but this
can partially be explained by the contribution of satellite galaxies, which are notoriously too
passive in semi-analytic models (e.g. Weinmann et al. 2006; 2011b). Models also fail to 
reproduce the anti-correlation between specific star formation rates (sSFR) and 
stellar mass (Somerville et al. 2008; Firmani, Avila-Reese \&
Rodr\'{i}guez-Puebla 2010). Finally,
the evolution of specific star formation rates (sSFR) in models seems to be inaccurate, with the sSFR too low
at z $<$ 2 (e.g. Daddi et al. 2007; Damen et al. 2009) and 
too high at $z>3$ (e.g. Bouch\'{e} et al. 2010; Weinmann et al. 2011a).

Hydrodynamical simulations of cosmological volumes today usually employ what is perhaps best
described as 'star formation-driven
galactic superwind feedback'. Two 
kinds of galactic superwinds (hereafter GSW) are commonly employed\footnote{We do not discuss in this paper high resolution hydrodynamical simulations of individual systems. We note
that these often have serious
problems in reproducing galaxy properties too (e.g. Guo et al. 2010; Scannapieco et al. 2012; Avila-Reese et al. 2011; but see also Brook et al. 2012) and
in addition it is not clear how to extrapolate their findings to the
overall galaxy population properties.}. 
The conventional approach is to use a fixed fraction of the energy
liberated by stellar feedback to drive winds with a constant wind speed and a constant mass loading.
Examples include the smoothed particle hydrodynamics (SPH) simulations
by Springel \& Hernquist (2003) and
Crain et al. (2009). 
These simulations fail to reproduce the stellar
mass function at $z=0$ (Crain et al. 2009). An alternative form of
 GSW feedback, based on momentum-conserving processes, has been proposed by Oppenheimer \& Dav\'{e} (2006), Oppenheimer \& Dav\'{e} (2008), Dav\'{e} et al. (2011a, b).
This scheme successfully
reproduces the low mass end of the 
stellar mass function and several other key properties of the observed galaxy population (e.g. Oppenheimer et al. 2010;
Dav\'{e} et al. 2011a, b). 
Interestingly, these momentum-driven wind models seem to suffer from 
similar problems as the semi-analytic models mentioned above
regarding the evolution of the mass function and the star formation rates (Dav\'{e} 2008; Dav\'{e} et al. 2011a).

We therefore conclude that models with star-formation
driven feedback as employed in most SPH simulations do not
reproduce the $z=0$ stellar mass function; models including different
feedback prescriptions, which either follow a scaling according to momentum-conservation, 
or the scaling usually used by semi-analytic models,
do manage to reproduce the low mass end of the $z=0$ stellar mass function, but fail in several other 
key aspects.

It is tempting to
infer from the discrepancies between models and observations that an
unknown process suppresses star formation in low mass haloes
at early times, potentially mitigating the need for strong
feedback at later epochs, like for example very inefficient high-$z$ star formation
(Krumholz \& Dekel 2012), preheating (Mo et al. 2005) or warm 
dark matter (as discussed in Fontanot et al. 2009).
Before continuing to explore  these options, it is appropriate to step back for a moment and
formulate more clearly what the problems of current \emph{models that broadly reproduce the stellar mass function at z=0,}
are, and how they relate to one another. To this end, we compare key galaxy properties in observations and
several state-of-the-art galaxy formation models in this work. We note
that most of the problems we described above become more severe towards lower stellar
masses. It is therefore useful to consider the lowest stellar mass bin
for which reasonably complete observational data and well-resolved
model results are  available. We choose the mass bin
log($M_{\rm
  star}/(h^{-2}M_{\odot}$))=9 - 9.5, or log($M_{\rm
  star}/M_{\odot}$)=9.27 - 9.77, which is
  the lowest stellar mass bin
where (i) robust estimates of stellar ages and star formation
rates for SDSS galaxies are still available for a significant number of
galaxies and (ii)
where galaxies are still resolved well enough in the models we
  use.\footnote{Galaxies of these masses
consist of $\sim$ 75 - 240 star particles in the simulations of Dav\'{e} et al. (2011a), and of $\sim$ 170 - 550 star particles 
in the GIMIC simulations.
Also, this is the mass where the semi-analytic model of De Lucia \&
  Blaizot (2007) is still resolution-converged between the
  Millennium-I and Millennium-II simulations (Guo et al. 2011).}

The failure of galaxy formation models to reproduce the observed
number density evolution of low mass galaxies is the key problem 
that we will investigate in this paper. In section \ref{sec:numden}, we
outline this fundamental discrepancy and  its relation to the 
number density evolution of dark matter haloes.

To explore the underlying causes for this problem and to find independent evidence for it, we then employ two
  different, complementary approaches.
In our first approach (Section \ref{sec:approachA}), we examine
  the specific star formation rates and luminosity-weighted ages of low mass central galaxies 
in the stellar mass bin given above in both observational data and
recent models.
For this, we use low-redshift observations
from SDSS; two semi-analytic models with different resolution 
and different prescriptions for astrophysical processes; the three SPH models
presented in Dav\'{e} et al. (2011a, b), of which 
one includes momentum-driven winds; and
the SPH simulation of Crain et al. (2009). For this part of the paper, we focus on central galaxies to isolate potential problems  in their
intrinsic evolution from those related to environment (that should mostly
affect satellite galaxies). We find a subpopulation of young and
  highly star forming galaxies in the observations that is absent in
  the models and that becomes more abundant towards lower masses, which is likely 
related to the problem in the number density/mass function evolution.

We adopt a more holistic approach in the second part of the
  paper (Section \ref{sec:toy}), where we construct a toy model that predicts the $z=0$ stellar
  mass function and galaxy number density given (i) the observed $z=1$ mass function and (ii) the
  observed  specific star formation as a function of stellar mass and
  redshift. With the help of this toy  model, we  demonstrate that the
  slow late evolution in the number density of low mass galaxies
  predicted by 
  the models is a consequence of an
incorrect relation between sSFR and stellar mass. This, in turn, may have its roots
in the growth rate of dark matter haloes, which scales very similarly
with halo mass and time like
the galaxy growth rate predicted by the models.

All quantities are quoted for $h$=0.73. The Guo et al. (2011) and GIMIC models are
 based on a WMAP1 cosmology (Spergel et al. 2003), the Wang et al. (2008) model on a WMAP3 cosmology
(Spergel et al. 2007), and the Dav\'{e} et al. (2011a,b) models
on a WMAP5 cosmology (Hinshaw et al. 2009). We convert redshift to lookback
time
using a WMAP1 cosmology.

\section{Data}
\label{sec:method}
\subsection{Observations at z=0}
All $z$=0 observations used in this paper are based on the SDSS DR4 (Adelman-McCarthy et al. 2006) and DR7 (Abazajian et al. 2009), using
two different samples.
The first  sample consists of the 16961 central galaxies in the Yang
et al. (2007) group catalogue with stellar masses
  $\log(M_{\rm star}/M_{\odot})$=9.27 - 9.77. Stellar masses are determined from fits to the photometry (see below).
 In most of what follows, we
restrict our analysis to the subset of galaxies with high-fidelity spectra (signal-to-noise $S/N>20$).
 This reduces our sample to 1630 galaxies.
Ages and metallicities from Gallazzi et al. (2005) are available for
9486 galaxies in the full sample, and for 1292 in the $S/N>20$ sample.

Our second sample consists of the 14719 central galaxies in the Yang et al. (2007) sample with stellar masses
 $\log(M_{\rm star}/M_{\odot})$=9.27 - 9.77 according to the Mendel et al. (in prep.) stellar mass estimates. Of those,
 709 have $S/N >$ 20. We note that the masses of Mendel et al. are higher 
than the Kauffmann et al. (2003) masses by on average about 0.15 dex, meaning that this second 
sample in effect consists of galaxies with slightly lower mass than the
first. We use the first sample everywhere except when
 using the Mendel et al. 
stellar age and metallicity estimates.

To correct for Malmquist bias, we weight observational results by 1/V$_{\rm max}$, with V$_{\rm max}$ the maximum volume out to which a given
galaxy can still be observed given the apparent magnitude limit of the survey.

\subsubsection{Yang et al. group catalogue}
We use the DR4 group catalogue\footnote{Publicly available at\\
\texttt{http://www.astro.umass.edu/∼xhyang/Group.html}}
 described in more detail by
Yang et al. (2007), and more specifically the sample 2 as described
by van den Bosch et al. (2008). The group catalogue
has been constructed by applying the halo-based group finder of
Yang et al. (2005) to the New York University Value-Added Galaxy Catalogue
(NYU-VAGC; Blanton et al. 2005). From this catalogue, Yang
et al. (2007) selected all galaxies in the Main Galaxy Sample with
an extinction-corrected apparent magnitude brighter than $m_{r}$ = 18,
with redshift in the range 0.01 $< z <$ 0.20 and with a redshift completeness
$C_{z} >$ 0.7. Group masses are derived from the summed stellar mass of
the galaxies in the group, with stellar mass 
estimates obtained according to Bell et al. (2003).

\subsubsection{MPA data}
We make use of the DR4 and DR7 SDSS data catalogues\footnote{Publicly available at
\texttt{http://www.mpa-garching.mpg.de/SDSS/}} 
from MPA/JHU, to obtain estimates for stellar masses, specific star formation rates, 
metallicities, 
stellar ages, and dust attenuations. We use the method updated for DR7
to calculate stellar masses and star formation rates, and the DR4 versions
for metallicities, stellar ages and dust. 
Stellar masses are 
estimated using fits to the photometry
and are similar to the estimates
from Kauffmann et al. (2003). They are based on a Kroupa IMF.
 Estimates of the aperture-corrected
specific star formation rates are based
on Brinchmann et al. (2004), with several modifications
regarding the treatment of dust attenuation and aperture corrections,
as detailed on the MPA webpage.

 Luminosity-weighted metallicities and luminosity-weighted stellar ages are obtained from Gallazzi et al. (2005).
To obtain estimates of dust attenuation, we use the $z$-band attenuation
by Kauffmann et al. (2003), which has been derived by comparing 
the fibre magnitudes with those computed using synthetic Bruzual \&
Charlot (2003) spectral energy distributions (SED)
that fit the fiber spectrum best.
We have converted this estimate into a $g$ and $r$-band attenuation using the Charlot \& Fall (2000) law. 
Dust attenuation can only be estimated within the fibre; we have, however, checked that there is no trend
of dust attenuation with redshift and thus with the fraction of the
  galaxy covered by the fibre. This enables us to apply the
dust attenuation measured within the fibre to the entire galaxy.

\subsubsection{Mendel et al. data}
We   obtain    alternative   estimates   for    SSP-equivalent   ages and
metallicities  from Mendel  et  al. (in  prep.). These estimates
are based on the Maraston (2005) stellar population models, and thus
complement the quantities estimated for our primary sample
based on the Bruzual \& Charlot (2003) stellar population models.
Briefly,  Mendel et al.  use the  SSP models  of Thomas,  Maraston \&
Johansson (2011) to interpret measured Lick line strengths in terms of
the luminosity-weighted age,  metallicity, and alpha-element abundance.
  Models  are fit via a grid  search using an adaptation
of the  multi-index chi-squared minimisation technique  discussed by Proctor
et  al.  (2004;  see  also   Thomas  et  al.  2010)  and  19  spectral
indices\footnote{Mendel  et al.  exclude Ca4227,  G4300,  Fe5792, NaD,
  TiO$_1$ and  TiO2 based  on the relatively  poor calibration  shown in
  figures 2,  3 and 4 of  Thomas, Maraston and  Johansson (2011)}.  In
instances  where data  are deemed  to be  poorly fit  by the  models, a
clipping procedure  is used  to remove the  index that results  in the
largest global reduction in chi$^2$. This procedure is iterated until a
good fit is obtained.   Relative to a simple sigma-clipping technique,
the method described  above naturally results in the  fewest number of
index removals to  reach an acceptable fit.  In  addition, it makes no
assumptions about the  relationship between the best fit  at any given
iteration and the final fit, and is therefore less likely to be biased
by single deviant indices.  Final  values of age, [Z/H] and [alpha/Fe]
are determined from the marginalised likelihood for each parameter.
Colours in the Mendel et al. sample are based on the updated photometry
of Simard et al. (2011), while stellar masses are derived in the
  same way as for the MPA data (see above), but with updated photometry, resulting in slightly 
higher masses.

\subsubsection{UV specific star formation rates}
To obtain an alternative estimate
for the sSFR, we use the UV star formation rates as obtained by McGee et al. (2011). 
These are available for 2194 galaxies in our sample.

\subsection{Semi-analytic models}
In order to model the evolution of galaxies, 
semi-analytic  models (SAMs) apply  analytic  recipes, describing the behaviour of the baryonic component, to  dark 
matter merger
trees 
(e.g. Kauffmann et al. 1993; Cole  et al. 2000).
We consider two semi-analytic models in this paper, namely Wang et al. (2008)
and Guo et al. (2011). 

Wang et al. (2008) is a variant of the De Lucia \& Blaizot (2007) 
model that has been adapted to a WMAP3 cosmology. It is run on top of a dark matter simulation with the resolution
of the Millennium Simulation (hereafter MS) with a dark matter particle mass
of $1.18 \times 10^{9} {\rm M_{\odot}}$ but in a volume that is a factor of 64 smaller than 
for the MS (i.e.
in a box with side 171 Mpc). Several parameters of the De Lucia \& Blaizot (2007) 
model have been
changed to adapt it to a WMAP3 cosmology, as described in detail in 
Wang et al. (2008).  The Wang et al. (2008) SAM has been tuned
to fit the $r$-band luminosity function at $z=0$. 

 Guo et al. (2011) present the first semi-analytic model
that has been applied to the Millennium-II (hereafter MS-II) high resolution simulation with 
a box of side 137 Mpc and a particle mass of $9.45 \times 10^{6} {\rm M_{\odot}}$ 
(Boylan-Kolchin et al. 2009). It is also based on De Lucia \& Blaizot (2007),
but
has been modified in several  aspects. In particular, the
efficiency  of  star-formation driven  feedback for low mass galaxies was  increased
considerably in  order to  fit the  low mass end  of the  stellar mass
function. The Guo et al. (2011) SAM reproduces the stellar mass and luminosity 
functions at $z=0$.

The production of metals in models is assumed to be instantaneous, and metals are
immediately fully mixed with the pre-existing cold gas. Metals are assumed to be transferred into the hot
and ejected gas phase, and reincorporated into the cold gas, in proportion to the gas itself
 (see De Lucia et al. 2004 for a detailed description). 
Luminosities are calculated according 
to Bruzual \& Charlot (2003). Both models use a Chabrier IMF and a slab dust model as described in De Lucia \& Blaizot (2007), with Guo et al. (2011) additionally allowing for a redshift evolution 
in the dust model, which should not affect any of our results. Luminosity-weighted ages 
for the Wang et al. (2008) model are calculated
in the $V$-band, following De Lucia et al. (2006).

\subsection{Hydrodynamical simulations}
Following the evolution of both dark matter and baryonic
particles in three dimensions, hydrodynamical simulation self-consistently trace the flow of baryonic matter into haloes. For
star formation and feedback, so-called 'subgrid recipes' 
are invoked, 
which vary between different simulations, and 
 have a strong impact
on the predicted galaxy properties (e.g. Schaye et al. 2010). Here, we use two 
state-of-the-art SPH simulations of cosmological volumes, the
Dav\'{e} et al. (2011a, b) simulations and the GIMIC simulations (Crain et al. 2009).
The simulations are complementary; whilst the GIMIC simulations
have
a slightly higher resolution and employ
a more standard stellar feedback prescription, the
Dav\'{e} et al. (2011a, b) momentum-driven wind simulation is
currently the only hydrodynamical simulation that accurately reproduces
the low-mass end of the stellar mass function in a cosmological volume.

\subsubsection{No winds, constant winds and momentum-driven winds simulations}
We use  three different variants of the SPH simulation that was introduced
by Oppenheimer et al. (2010) and explored in more detail by Dav\'{e} et al. (2011a, b).
These simulations have been run with an 
 extended version of the Gadget-2 N-body + SPH code (Springel 2005; Oppenheimer \& Dav\'{e} 2008) in a box with side 48 $h^{-1}$ Mpc, with 
$384^3$ dark matter and $384^3$ gas particle. The gas particles mass is $3.6 \times 10^{7} M_{\odot}$, 
with star particles on average half as massive and
dark matter particles 
having a mass of $1.8 \times 10^{8} M_{\odot}$. We make use of their
``no winds'', ``constant winds'', and ``momentum-driven winds'' models in this work, 
which are referred to as ``nw'', ``cw'', and ``vzw'' in what follows, to follow
the nomenclature of the original papers.
These models differ in their treatment of  feedback. 

In the nw model, feedback energy is imparted via (inefficient) thermal
heating of the ISM, 
using the Springel \& Hernquist (2003) subgrid two-phase recipe. In both the cw and the vzw model, 
kinetic feedback
is added by explicitly kicking individual particles, causing GSW feedback.
The cw and vzw models differ in their wind velocity, and
the mass of gas that is accelerated per unit stellar mass formed (the
'mass-loading factor').
In the cw model,
particles are kicked with initial velocities of 686 km/s, and constant mass-loading of 2,
  corresponding to 95 \% of Type II SN energy being converted to
  outflows if all stars with masses above 10 $M_{\odot}$ 
end their lives as a supernovae
  (Oppenheimer et al. 2012).
In the vzw model, the wind
velocity is proportional to the velocity dispersion of the galaxy, and the 
mass-loading is inversely proportional to it. Such a scaling is expected if 
the energy source of the winds is momentum transfer from UV photons coming from 
massive stars and might naturally occur from a combination of
different feedback mechanisms (e.g. Hopkins et al. 2012). In terms of energetics, the vzw model has more modest requirements
than the cw model. For instance, a
$M_{\rm star}= 10^{10} M_{\odot}$ galaxy only uses 30 \% of the
available SN energy at $z=1$ and only 21 \% at $z=0$ for powering
winds (Oppenheimer et
al. 2012).

The vzw model was initially tuned to match CIV absorption in quasar
absorption line spectra at $z=2-5$ (Oppenheimer \& Dav\'{e} 2006).  It also
reproduces 
the present-day SMF below the knee of the mass
function, and several other important galaxy population properties
like the mass-metallicity relation at $z=2$
(Finlator \& Dav\'{e}  2008). Due to the
absence of AGN feedback, the model does not reproduce the high  mass
end of the stellar mass function, and the global star formation rate
at $z<1$. 

We use instantaneous SFR in what follows, which is calculated
from the instantaneous gas density. Luminosities are calculated according to Bruzual \& Charlot (2003), using
a Chabrier IMF. The simulations
account for metal enrichment from Type II and Type Ia supernovae and asymptotic giant branch
stars, and track four elements (C, O, Si, Fe) individually, as described
in Oppenheimer \& Dav\'{e} (2008). We approximate the stellar
metallicity of simulated galaxies with
\begin{equation}
Z_{\rm stellar} = (\rm{Fe} + 0.93 \times {O})/1.93,
\end{equation} 
where Fe and O is the iron and oxygen mass fraction of the stars, scaled to a solar value
of  0.001267 and 0.009618 (Anders \& Grevasse 1989).
Dust attenuation in these models is estimated according
to Finlator et al. (2006) in an empirical fashion, following the observed dust-metallicity relation.

\subsubsection{GIMIC simulations}
The \textsc{Galaxies-Intergalactic Medium Interaction Calculation} (GIMIC; Crain et al. 2009) simulations use
the Gadget-3 SPH and N-body code, with 
star formation,  stellar feedback, radiative cooling and chemodynamics as described 
in Schaye \& Dalla Vecchia (2008), Dalla Vecchia \& Schaye (2008), Wiersma, Schaye \& Smith (2009a) and
Wiersma et al. (2009b).  The 
star-formation driven feedback implementation is conceptually similar to the 
cw model described above, but with a mass loading  of 4, which is twice as high as in the cw model, and a wind
velocity of 600 km/s. This means that 80 \% of the SN energy is used
for powering winds, assuming that all stars with mass  above 6 $M_{\odot}$ go supernova.  While in the cw model, winds are temporarily hydrodynamically 
decoupled, this is not the case in the GIMIC simulations, the consequences of which are outlined in 
Dalla Vecchia \& Schaye (2008).
The choice of the mass loading factor is motivated by the desire to
produce a star formation 
rate density evolution broadly compatible with observations. The GIMIC simulations adopt the Chabrier IMF.

GIMIC re-simulates five environmentally-diverse
regions extracted from the MS simulation at a higher resolution. The regions enclose spheres with a radius of
20 $h^{-1}$Mpc. In what follows, we use the weighted mean result
  of the five regions.

We use the intermediate resolution realisation of GIMIC, in which gas
particles have a mass of 1.6 $\times 10^{7} M_{\odot}$ with the
  star particle approximately half as massive for most of their lives,
 if one takes into account
stellar recycling, and
a DM particle mass of 6.6 $\times 10^{7} h^{-1} M_{\odot}$. This corresponds to a 
twice as high mass resolution
than in the simulations of Dav\'{e} et al. (2011a,b). 

The GIMIC simulations have been shown to reproduce several properties of $L^\star$ galaxies, such as their X-ray to optical luminosity scaling relations (Crain et al. 2010), their stellar halo structure and dynamics (Font et al. 2011; McCarthy et al. 2012a) and their Tully-Fisher relation (McCarthy et al. 2012b).

\subsection{Warm dark matter-only simulations}
We also use data from two warm dark matter (hereafter WDM)-only simulations. These were run 
with PKDGRAV (Stadel 2001) in 
a box of side 90 Mpc/h, containing $400^{3}$ particles with a dark matter particle mass of $7.0 \times 10^8
h^{-1} M_{\odot}$. Simulations are based on a WMAP5 cosmology, with the power spectrum truncated 
as suggested by Viel et al. (2005), using the analytical expression in Macci\`{o} et al. (2012). We 
explore two different assumptions for the WDM mass: 2 keV, which
is the lower
limit of the warm dark matter mass 
consistent with constraints from the Lyman-alpha forest (e.g. Seljak et al. 2006) and 0.5 keV.
Haloes were identified using the spherical overdensity halo finder of
Macci\`{o} et al. (2008), imposing a minimum halo mass of 100
particles. At this halo mass, no compelling evidence of spurious
structure (see e.g. Wang \& White 2007) was found in the simulation
\section{The problem}
\label{sec:numden}
\begin{figure*}
\centerline{\psfig{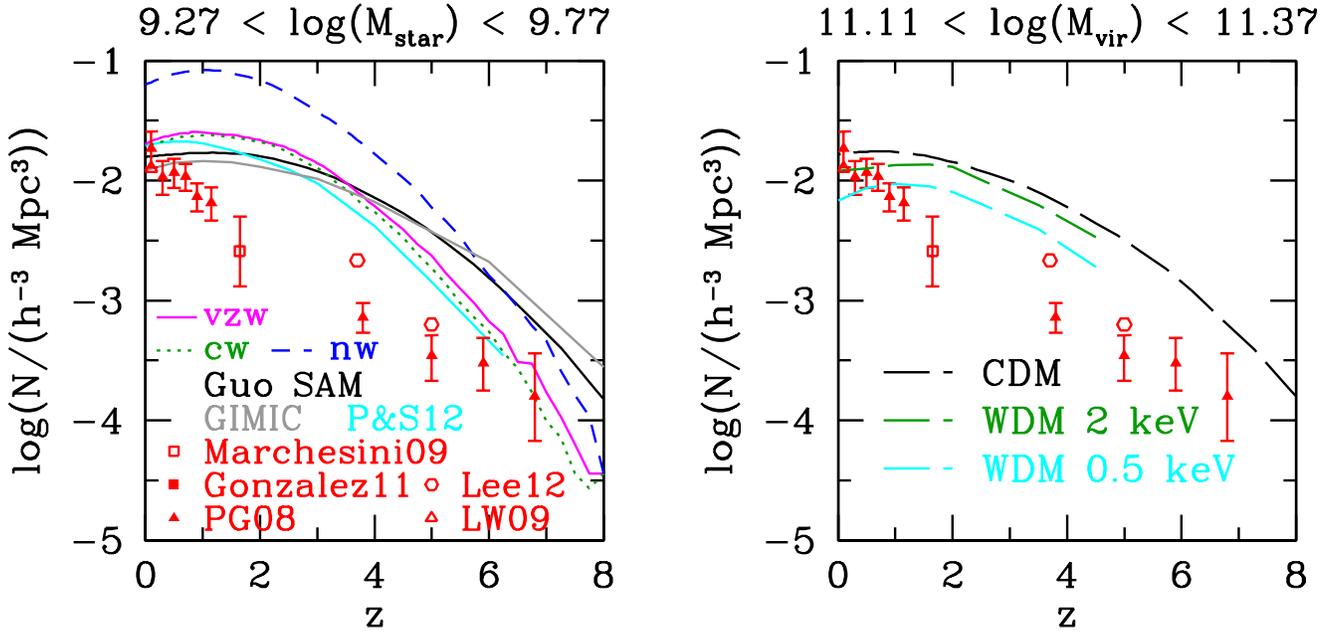}}
\caption{The evolution of the number density of galaxies with masses
$\log(M_{\rm star}/M_{\odot})$=9.27-9.77 as a function of redshift. 
Red symbols with errorbars denote observational results. 
In the left panel, various model predictions are shown.  Black lines denote results from the Guo et al. (2011) SAM,
magenta solid lines are for the vzw model, blue dashed lines for the nw model, 
green dotted lines for the cw model, grey lines show results from
GIMIC, cyan lines
show results from Puchwein \& Springel (2012).
In the right hand panel, we compare observations to the number
density evolution of haloes with masses $M_{\rm vir}=11.11 - 11.37$. Black dashed lines
show results from the MS-II simulation, dark green and cyan dashed lines show results from two different
WDM models, with a dark matter particle mass of 2 and 0.5 keV respectively.
}
\label{fig:evolution}
\end{figure*}

The central motivation of this study is the discrepancy in the evolution in the stellar mass function
between semi-analytic models and observations as found by Marchesini et al. (2009), Fontanot et al. (2009) and Guo et al. (2011).
We illustrate this discrepancy in Fig. \ref{fig:evolution}, using updated observational results and
results from the models described in the previous section,
showing that the same problem occurs in hydrodynamical simulations.

More specifically, we show the number density evolution of galaxies
in the stellar mass range 9.27 $<\log(M_{\rm star}/M_{\odot})<$ 9.77 versus redshift.
Black and red data points on both panels show observational results from 
P\'{e}rez-Gonz\'{a}lez et al. (2008), Li \& White (2009), Lee et al. (2012),
Gonz\'{a}lez et al. (2011) and Marchesini et al. (2009)\footnote{For
  P\'{e}rez-Gonz\'{a}lez et al., we use the datapoints and errors of the I-band MF. For Marchesini et al., integrate their Schechter
fit in the relevant range, but use errors as given for this mass bin in their Table 1. For Li \& White (2009)
and Lee et al. (2012), we integrate
the analytical stellar mass function.  Only datapoints within the completeness limits
given by those authors are used.}. All observed stellar masses
have been scaled to a Chabrier IMF.
The offset at $z=0.1$ between P\'{e}rez-Gonz\'{a}lez et al. (2008) and Li \& White (2009) is
likely due to cosmic variance affecting the P\'{e}rez-Gonz\'{a}lez et al. estimate.

In the left panel, we compare observational results with 
predictions from various galaxy formation models. Black solid lines
show results from the Guo et al. (2011) SAM. The solid magenta, dashed
blue, 
green dotted, and grey lines show results
from the vzw, nw, cw and GIMIC SPH simulations, respectively. On this plot alone, we
also include results from a SPH simulation featuring a new 'energy-driven variable wind' 
feedback scheme, recently suggested by Puchwein \& Springel (2012). These results are shown 
as the cyan line.
Clearly, there is a significant discrepancy between models and observations. For example, in the 
dataset of P\'{e}rez-Gonz\'{a}lez et al. (2008) the number of galaxies with 
$\log(M_{\rm star}/M_{\odot})=9.27-9.77$ increases by a factor of 2 from z=0.9 to z=0.1.
In the SAM of Guo et al. (2011), it \emph{decreases} by 7 \% in the
same redshift interval. The data point of Marchesini et al. (2009) at $z \sim 1.6$ is
lower than any model by a factor of $\sim$ 5. 
 More generally, the number of galaxies in the observations
is roughly inversely proportional to redshift over the entire redshift
range probed, while the models predict that the number density only
increases steeply from high redshift to $z=1-2$,
 and then flattens. Turning to the SPH simulations, 
the GIMIC simulation resembles the SAM most closely, while the cw and
vzw simulations show a steeper evolution, probably 
due to their lower mass resolution. The nw simulation, on the
other hand, strongly overpredicts the number density of galaxies at
$z=0$, which is not surprising given the absence of a strong feedback
mechanism in this model. Finally, the simulation by 
Puchwein \& Springel (2012) is slightly closer to observations at $z>0$, but 
a large discrepancy remains.

It is interesting to now consider the evolution of dark matter haloes that
are likely to host the galaxies we consider here. 
In the right panel of Fig. \ref{fig:evolution}, the black dashed lines show results for the evolution 
in the number density of dark matter haloes in MS-II, with $\log(M_{\rm vir}/M_{\odot})$\footnote{$M_{\rm vir}$ is defined
as the dark matter virial mass for centrals, and as the
virial mass just before infall for satellite galaxies.} =11.11 -
11.37, which typically host the galaxies we consider at $z=0$ in
  the SAM. The resulting functional form is very similar to what has
been found previously for dark matter-only simulations (Luki\'{c} et al. 2007).
Clearly, the number density evolution of dark matter haloes
of this mass is also very similar
to that of the galaxies in the SAM (compare the black line in the left
hand panel
with the black dashed line in the right hand panel).
This indicates
that the relation between stellar mass and host halo mass barely evolves in the models: a galaxy with mass $\log(M_{\rm star}/M_{\odot})=9.27-9.77$
resides in a host halo with a very similar mass up to high redshifts
(see also Sec. \ref{sec:recon}). 
If we compare the dark matter halo number density to the 
observational results, observations seem to require that the relation  between
stellar mass and halo mass evolves strongly in a  $\Lambda$CDM cosmology. Haloes of fixed mass must host
galaxies with lower stellar mass at higher redshift (as also found
 e.g. by Moster et al. 2010, 2012; Behroozi et al. 2010;
Yang et al. 2011; Zehavi et al. 2012).

Given that it has been suggested that WDM might help to solve the problem,
we also plot the number density evolution of haloes in this mass range in two 
WDM simulations with a dark matter particle mass of 2 keV and 0.5 keV.
We find that while the normalisation changes slightly, the way the number density
of haloes evolves is essentially unchanged with respect to CDM, indicating
that the problem would likely also be present in a WDM Universe.

In summary, we confirm that the number density of low-mass galaxies evolves incorrectly in SAMs
and verify that the problem also exists in SPH simulations.  We suggest that, in both models, the
growth of low-mass galaxies traces the growth of their host halos too closely, and that the mass 
of their host halos should increase to higher redshifts as also found in recent clustering and abundance
matching studies (e.g. Yang et al. 2012).

\begin{figure*}
\centerline{\psfig{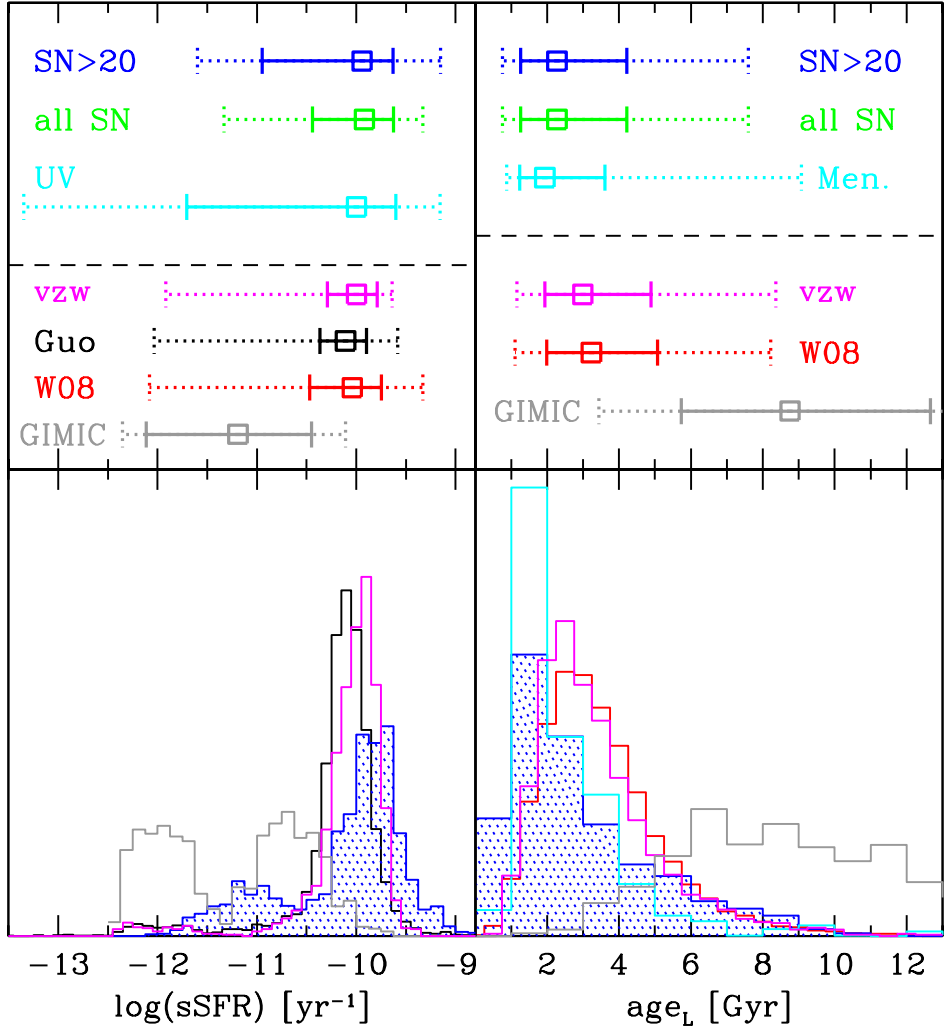}}
\caption{Model and observational results for specific star formation rates and luminosity-weighted stellar ages in the stellar mass bin $\log(M_{\rm star}/M_{\odot})=9.27-9.77$.
Top panels
show the median values (empty squares), the range within which 68 \% of the values lie (solid errorbars) and
the range within which 95 \% of the values lie (dotted errorbars). Bottom panels show the full distributions
for several selected datasets. The colour coding is as follows.
Left panel:
Blue and green: updated 
Brinchmann et al. (2004) sSFR, S/N$>$20 and the full sample. 
Cyan: UV sSFR from McGee et al. (2011). Magenta: vzw simulation. Black: Guo et al. SAM. Red: Wang et al. (2008) SAM. 
Grey: GIMIC simulation.
Right panel: Blue and green: Gallazzi et al. (2005) luminosity-weighted
ages for the $S/N>20$ and the full sample. Cyan: Mendel et al. sample. Magenta: vzw simulation. Red: Wang et al. (2008)
SAM. Grey: GIMIC.
   }
\label{fig:sfr}
\end{figure*}

\begin{figure*}
\centerline{\psfig{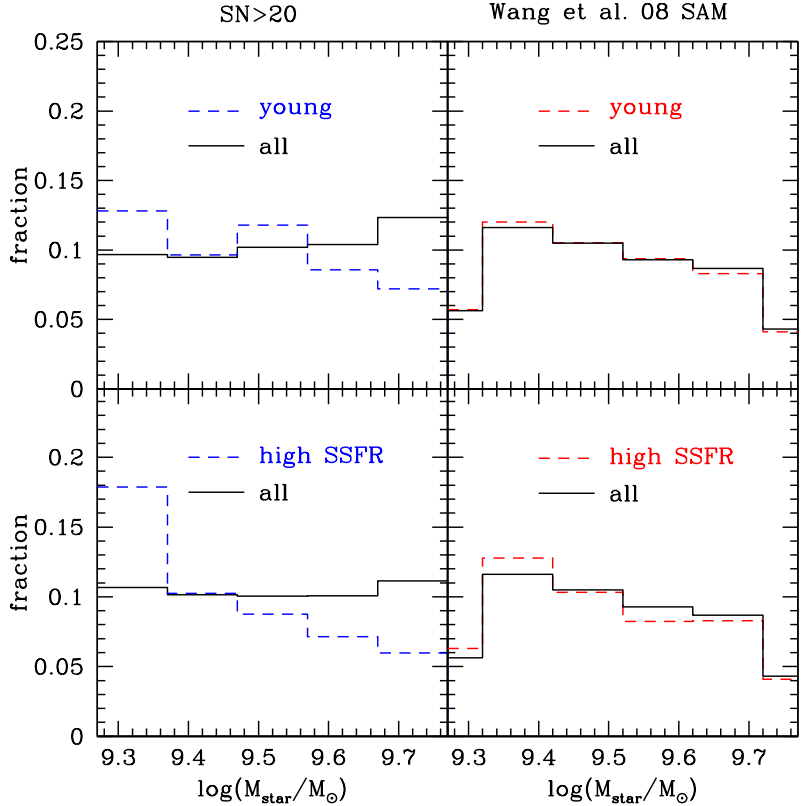}}
\caption{Stellar mass distributions in the observational sample and
  the W08 SAM in the stellar mass bin $\log(M_{\rm star}/M_{\odot})=9.27-9.77$. 
The black solid line is for the full sample, the coloured dashed lines are for the subpopulation of galaxies
with ages below 2 Gyr (top panels) and $\log(\rm{sSFR})>$-9.5 (bottom panels). Clearly, the strongly 
star forming and young galaxies tend to have low masses in the observations, while there is nearly 
no such trend in the models. The full sample is slightly different in the top and bottom left panel, since age estimates are not available for all galaxies. }
\label{fig:young}
\end{figure*}

\section{Approach I -- Low mass galaxy properties}
\label{sec:approachA}
We proceed to  compare specific star formation rates  and stellar ages
for galaxies  with masses $\log(M_{\rm  star}/M_{\odot})$=9.27-9.77 in
the observations and models. We only focus on central galaxies, since
  satellite galaxies  are not correctly reproduced  in current models
(e.g.  Weinmann et al.  2011b). 

While in the observations and in the models
of Dav\'{e} et al. (2011), satellites seem to be a subdominant population at 
these stellar masses, this is not the case for the semi-analytic models. 
The
Guo et al. (2011) SAM predicts that as many 
as  50 \%  of the  galaxies at $z$=0  in the  stellar mass
range  we consider  are satellites.

 We   do not include the  constant-wind  and  no-wind simulations  by
Dav\'{e} et  al. (2011a, b) in the following comparisons, since  the constant-wind
simulation is similar to GIMIC, and the no-wind simulation is very far
off the observations at $z$=0.

\subsection{Low redshift results}
\label{sec:z0}

\subsubsection{Specific star formation rate}
\label{sec:ssfr}

In Fig. \ref{fig:sfr}, left panels, we show the distribution of
specific star formation rates in our various datasets, including models
  and observations.
The top left panel shows the median sSFR as an empty square, the range within which 68 \% of galaxy sSFR lie as
solid errorbars, and the range encompassing 95 \% of the sSFR as
dotted errorbar. The bottom panel
shows the full distribution for several selected subsets. Model galaxies
with a SFR of zero are assigned a random value between log(sSFR)=-11.6 and -12.4.
  The sSFR from UV
and from the updated Brinchmann et al. (2004) method are in good
agreement, except for an extended tail of low sSFR galaxies in
  the UV, which corresponds to the UV non-detections.

The agreement between the median log(sSFR/yr) in the 
observations ($\sim -9.95$ ) and the vzw model ($\sim -10.0$) is
surprisingly good, given the large discrepancy between observed and model sSFR at low masses found by Dav\'{e} et al. (2011a).
The reason for this difference appears to be that Dav\'{e} et al. (2011) used
 observations by Salim et al. (2007) that were restricted to 
a star-forming sample of galaxies to compare to their model galaxies.
The median log(sSFR/yr) in the Guo et al. (2011) model is slightly
lower ($\sim -10.1$), but also in relatively good agreement with
  observations, in contrast to the substantial offset found in previous work
  (Fontanot et al. 2009; Guo et al. 2011). This difference, in turn, is
due to the exclusion of satellite galaxies in our comparison, whose
properties are likely incorrect in SAMs.

Crucially however, the models seem to miss a tail of high star formation rates that
are present in the observations. While 9 \% of galaxies in the S/N$>$ 20 sample have $\log$(sSFR)$>-9.5$, this
is only the case for 1.5 \% of galaxies in the Guo et al. (2011) model, and for 0.4 \% of galaxies
in the vzw model.  We note that if these galaxies have formed stars at the current
rate or higher in the past, they  have formed entirely in less than 3 Gyr, i.e. since $z$=0.3.
About 2.5 \% of galaxies even have log(sSFR) in excess of -9.2, meaning they could  have formed
all their stars within the last 1.6 Gyr, or since z=0.15.
The fact that we find such highly star forming galaxies independent of whether 
the UV sSFR or the Brinchmann et al. (2004) sSFR
is used indicates that the discrepancy with models is robust. 
We have checked that a similar picture is seen at $z=1$ in the ROLES data by Gilbank et al. (2011)
who use the O[II] line to estimate star formation. At $z=1$, $\sim$ 15 \% of galaxies have doubling times of less than 1 Gyr in the observations, while this is the case for less than 2 \% in the Guo et al. (2011) model.

At the other end of the distribution, we find 14 \% of galaxies in the S/N$>$20 sample with 
$\log$(sSFR)$<$-11. Only 6 \% of galaxies in the Guo et
al. sample have such low star formation rates, and less than 4 \% in the vzw sample. 
This difference could be due to (i) contamination of the observed sample by 
satellite galaxies or (ii) a real quiescent population of central galaxies that is
more abundant than in the model.
(i) seems an unlikely explanation, as the contamination of the central sample by satellites is estimated to be 
only around 3 \% (Weinmann et al. 2009).
Our result might thus be in tentative 
agreement with the population of red, isolated, faint central galaxies
in the SDSS that seems to have no counterpart in semi-analytic
models (Wang et al. 2009).

We note that the GIMIC simulation strongly underpredicts star formation rates. While 
this might partially be because star-formation rates are not
resolution-converged in GIMIC, the discrepancy we find seems in line with
the usual problems of standard hydrodynamical simulations (e.g. Avila-Reese et al. 2011). In general,
galaxies in those simulations form in an early strong burst of star formation (Scannapieco et al. 2012), leading to strong
feedback that makes the galaxies almost passive by the present day
(see Section \ref{sec:sfh}). Since GIMIC 
does reproduce the evolution of the cosmic star formation rate
density, star
formation in this simulation  is overly 
concentrated in massive galaxies at low redshift (see Fig. 5 of Crain et al. 2009).

As mentioned before, the good agreement between observations and the
semi-analytical model is due to the exclusion of satellites. If we include
satellites, the median sSFR of SDSS galaxies only decreases slightly compared
to the full sample, to log(sSFR)=-10.05. The median sSFR of galaxies in the Guo et al. 
sample, however, goes down to log(sSFR)=-10.43. Also, the passive fraction (with 
log(sSFR) $<$ -11) in the 
observations increases only to 24 \%, while it reaches 42 \% in the Guo et al. model. 
Including satellites  has a much more moderate effect in the
 vzw simulation; it
changes the median to log(sSFR)=-10.04, in excellent agreement with observations, 
and increases the fraction of passive galaxies only to 14 \%, which is still lower than the
passive fraction of the full observed sample, indicating that the satellites in the vzw model
are in fact quenched too little.

We conclude that there seems to be insufficient diversity in the star formation rates of
model galaxies. Part of the reason for
this could be that star formation in the models is not sufficiently bursty.
Also, the median sSFR is slightly too low in models. Both of
  these problems may be
  related to the weak evolution in the number density of model
  galaxies at late times.

\subsubsection{Luminosity-weighted Stellar Ages}
In Fig. \ref{fig:sfr}, right panels, we show
luminosity-weighted\footnote{Both for models and observations, the luminosity-weighting
refers to dust-free luminosities.} ages from Gallazzi et al. (2005, blue
lines), and  from Mendel et al. (cyan lines). They are compared
to $r$-band luminosity weighted ages from the vzw-model (magenta lines) 
and  $V$-band luminosity-weighted ages from Wang et al. (red lines). 
We convolve all model results with a Gaussian distribution with $\sigma$=0.15 in logarithmic age
which is the mean 68 \% confidence range as indicated in Gallazzi et al. (2005) for
our stellar mass bin. 

Model and observed distributions are clearly different.
In the $S/N>20$ sample of Gallazzi et al. (2005), 
more than 40 \%\footnote{Including galaxies with lower S/N increases the observed
fraction of young galaxies even more, to over 50 \%. We have checked that only using galaxies
at redshifts at $z<0.03$, and no volume-weighting, does not change the observational results.} of galaxies have ages below 2 Gyr, while
this is the case for only 16 \% in the Wang et al. (2008) sample.  
The luminosity-weighted ages of Mendel et al. and of Gallazzi et al. (2005) are in good agreement, despite
being based on two different SSP models (Maraston 2005
vs. Bruzual \& Charlot 2003). This 
indicates that the observational result is robust.

These results  indicate that the models not only lack the 
subpopulation of  galaxies currently having high star formation rates
 (see Section \ref{sec:ssfr}), but also predict too little star formation in the last 2-3 Gyr. We have checked that while
sSFR and age are broadly anti-correlated in the observations as expected,
many of the galaxies with 
young ages do not have particularly high current sSFR. 
A similar problem, but regarding 
ages weighted according to stellar mass and not according to
  luminosity (which are more uncertain, see Gallazzi et al. 2008)
has been found by Somerville et al. (2008), Fontanot et al. (2009) and
Pasquali et al. (2010).

\subsubsection{What are the young and star forming galaxies?}
\label{sec:young}
We have pointed out that the fraction of young (ages below 2 Gyr) and very active (log(sSFR)$>$-9.5)
galaxies is higher in the observations than in the models.
In Fig. \ref{fig:young} we show the distribution in stellar mass for the young and active galaxies
compared to the full sample both for the $S/N>20$ observations, and for the Wang et al. (2008) model. We use
a binning of 0.1 dex, which is half the expected error in stellar mass according to Gallazzi et al. (2005). 
Using a larger
binning does not change the basic trends that these figures show: In the observations, lower mass
galaxies are on average younger and have higher sSFR. \emph{This indicates that it is especially the galaxies just entering
our stellar mass bin which are too passive and too old in the model. From this it follows that the
rate of galaxies entering the mass bin is probably too low, which may explain the 
missing evolution in the number density we found in Fig. \ref{fig:evolution}.}

\section{Approach II -- Toy model}
\label{sec:toy}
In our second approach to understanding the number density evolution of
low mass galaxies, we employ a simple toy model to check if the
observed evolution of the stellar mass function and the sSFR-stellar
mass relation can be reconciled. 
We find that this is the case if we adopt a relatively shallow slope
in the sSFR-stellar mass relation, as found by several observational
studies.

We start with the analytical fit to the stellar mass function at $z$=0.9, as obtained by
P\'{e}rez-Gonz\'{a}lez et al. (2009), cut off at
 $M_{\rm star}=10^{7} M_{\odot}$. This stellar mass function 
is then evolved up to the present day, assuming that all galaxies 
follow the same relation 
sSFR($M_{\rm star}, z$) and that 40 \% of newly formed stars are immediately
returned to the ISM, according to a Chabrier IMF.   The final stellar mass function is then compared
with the $z$=0 stellar mass function obtained
by Li \& White (2009) (including the       correction by Guo et
al. 2010).  Both the
P\'{e}rez-Gonz\'{a}lez et al. (2009) and Li \& White (2009) MF have been scaled to a
Chabrier IMF.

Our toy model is based on the fact that the evolution of the mass function can be described by a simple continuity 
equation, as explained in more detail in Drory \& Alvarez (2008). As
an input to this continuity equation, we need the
average relation between sSFR and stellar mass for the full population of galaxies. The scatter around that relation, and the fraction of passive galaxies, 
however, do not need to be known.  
Similar approaches have been used by Bell et al. (2007), and by Peng et al. (2010). Like these models, our toy model requires
extrapolation of the stellar mass function and the sSFR-stellar mass relation below the observational limits.

\subsection{The observed sSFR-stellar mass relation}

Following Karim et al. (2011), we parameterize the relation between
star formation, redshift and mass as:
\begin{equation}
\log({\rm sSFR (M_{\rm stellar}, z)})=C + \beta\log(M_{\rm stellar}) + \alpha\log(1+z).
\label{eq:ssfr}
\end{equation}
$\alpha$ is usually found to be about 3-4.5 (e.g. Damen et al. 2009;
Karim et al. 2011; Fumagalli et al. 2012).
$\beta$, on the other hand, that parameterizes 
the correlation between sSFR and stellar mass, is usually found to be negative
in observational studies up to at least $z \sim 2$. Karim et al. (2011), for example, find
$\beta \sim$ -0.4 and $\beta \sim$ -0.7 for their full/star-forming
samples respectively, Drory \& Alvarez (2008), including an incompleteness correction
for galaxies with low SFR,  find
$\beta$ between -0.3 and -0.4, Noeske et al. (2007) find
$\beta \sim$ -0.3 for star-forming galaxies, and Whitaker et al. (2012) find
$\beta \sim$ -0.4 for their full sample. A less steep slope is advocated by Salim et al. (2007; $\beta$=-0.17
for low mass star forming galaxies).  Elbaz et al. (2007), Daddi et al. (2007) and Dunne et al. 
(2009) all find  $\beta$=-0.1, with the former two samples being restricted to star-forming galaxies, 
and the latter referring to a  $K$-band selected sample.

\begin{figure}
\centerline{\psfig{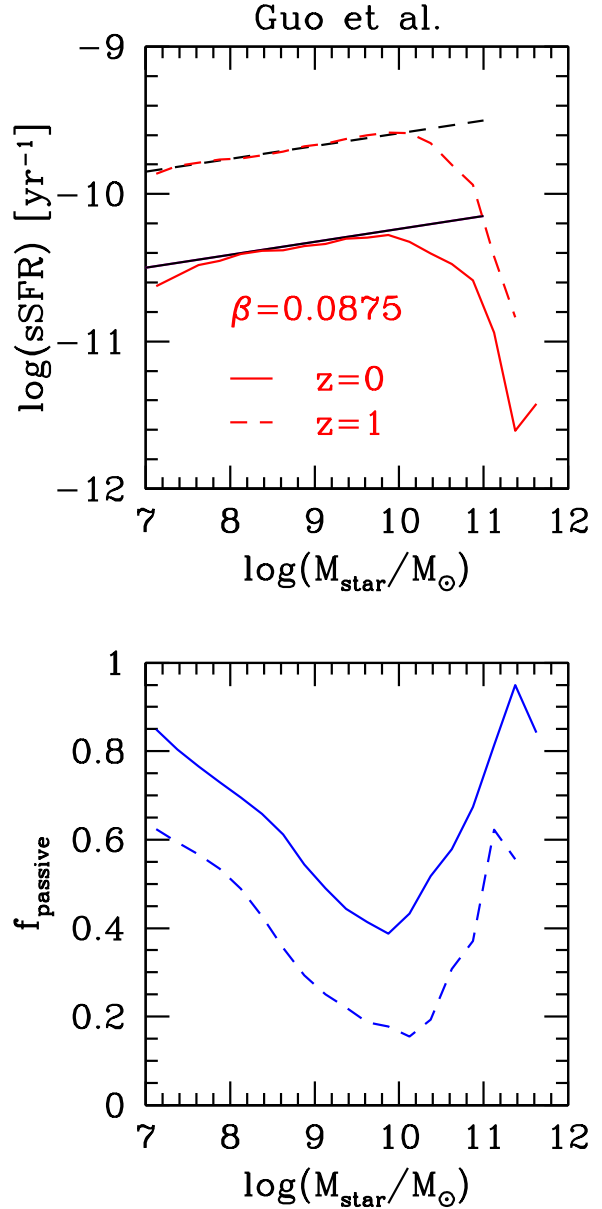}}
\caption{Mean sSFR (top panel, in red) and passive fractions (with
  log(sSFR)$<-11$, bottom panel, in blue) in the Guo et al. (2011) SAM at $z=0$ (solid lines) 
and $=1$ (dashed lines) as a function of stellar mass. 
The average relation between sSFR and mass in the SAM is almost perfectly fit by a 
positive slope of $\beta$=0.0875, while the evolution in redshift is fit by a factor of $(1+z)^{2.15}$ (black lines).
This is very close
 to the corresponding relations for dark matter haloes (Neistein \& Dekel 2008), 
but at odds with observational results.}
\label{fig:tilt}
\end{figure}

\subsection{The sSFR-stellar mass relation in models}

We check the same relation in the Guo et al. (2011) SAM in
  Fig. \ref{fig:tilt}, top panel, 
where we show the mean relation between log(sSFR) and $\log(M_{\rm
  star})$ at $z=0$ and $z=1$ in the SAM (including satellite and central galaxies), to 
which we fit a linear relation. We find that $\beta$=0.0875
and $\alpha$=2.15 provides a good fit 
at $\log(M_{\rm star}/M_{\odot})$=7-11 both at $z=0$ and $z=1$. If we
exclude satellite galaxies, $\alpha$ and $\beta$ remain virtually unchanged
(only the normalization of the sSFR shifts up).
Remarkably, both $\alpha$ and $\beta$ as found in the Guo et
al. (2011) SAM closely resemble the scaling expected for the specific
dark matter accretion rate, where $\beta \sim 0.1$ and $\alpha \sim
2.2$ (Neistein \& Dekel 2008). This similarity between sSFR and
specific dark matter accretion rate is interesting given the presence of
strong feedback in the SAM; in hydrodynamical simulations, the
baryonic
accretion rate already deviates from this scaling
(Faucher-Gigu\`{e}re et al. 2012).

In the lower panel of the same Figure, we show
the passive fraction of galaxies as a function of stellar mass in the
SAM for illustration. Clearly, the passive fraction strongly increases
towards lower stellar mass, and this is what is causing the positive 
correlation between stellar mass and sSFR. If we remove all passive galaxies
in the SAM, we find $\beta \sim 0$ at $z=1$ and $\beta \sim -0.1$ at
$z=0$. This means that the positive slope in the SAM comes from an 
overprediction of the number of passive galaxies in the model. 

We note that sSFR and stellar mass are even more strongly positively correlated in the vzw model, 
where
$\beta \sim 0.25$ at $\log(M_{\rm star}/M_{\odot})=9-10$ (see Dav\'{e} et al. 2011)\footnote{The reason for this is differential wind recycling where
high mass haloes are more efficient in reaccreting ejected mass (Oppenheimer et al. 2010;
Firmani, Avila-Reese \&
Rodr\'{i}guez-Puebla 2010).}.

\begin{figure*}
\centerline{\psfig{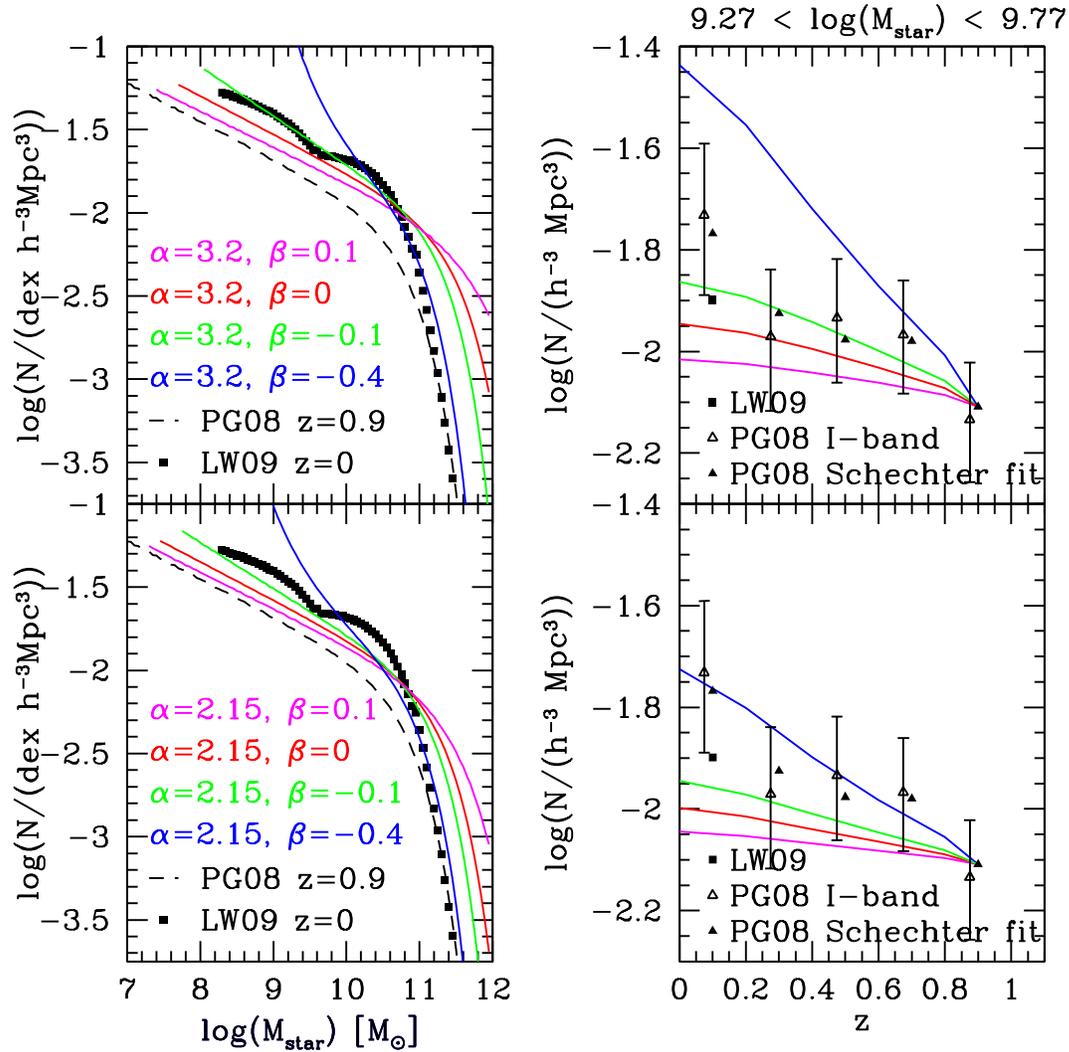}}
\caption{A toy model showing the evolution of the stellar mass function 
starting from the $z=0.9$ mass function by P\'{e}rez-Gonz\'{a}lez et al. (2008) 
(dashed line in the left panels) and assuming all galaxies follow the same sSFR($z$, $M_{\rm star}$)
with varying $\alpha$ and $\beta$, where $\alpha$ expresses the dependence of the 
sSFR on (1+$z$), while $\beta$ denotes the slope in the sSFR-stellar mass relation.
Top panels are for $\alpha=3.2$, bottom panels for $\alpha=2.15$.
In the left panels, we compare the toy model results to the observed evolution
of the stellar mass function. Solid black squares show the stellar mass function 
from Li \& White (2009) at $z=0$. In the right panels, we show a plot similar to Fig. \ref{fig:evolution}, truncated
at $z=1$. As we need to use the analytical stellar mass function at $z=0.9$, 
we also show the number density obtained from integrating the analytical 
Schechter form of the stellar mass function as obtained by P\'{e}rez-Gonz\'{a}lez et al. (2008) (points
without error bars). They are notably offset from the directly measured number densities, 
which indicates that the Schechter fit is not perfect at the masses we consider.}
\label{fig:toymodel}
\end{figure*}

\subsection{Results of the toy models}
In Fig. \ref{fig:toymodel}, we show 8 simple models with
varying $\alpha$ and $\beta$, and compare them with both the stellar mass function evolution
(left panels) and the evolution in the number densities (right panels) since $z=0.9$. In the top panels, 
we use $\alpha$=3.2, following observational results of Fumagalli et al. (2012),
 in the bottom panels we use $\alpha$=2.15, following the SAM. We vary $\beta$ 
between 0.1, 0, -0.1 and -0.4, and fix
 $C$ such that log(sSFR)=-9.95 at $\log(M_{\rm star}/M_{\odot})$=9.52, 
as we find in the SDSS\footnote{This is the median value for centrals alone, which is
good enough for our purposes here. The mean sSFR for centrals and satellites together, 
which one could argue should be used, is slightly higher, 
log(sSFR)=-9.89 at $\log(M_{\rm star}/M_{\odot})$=9.55.}. 
The results show that that the amount of evolution in the stellar mass function 
between $z=1$ and $z=0$ depends very strongly on $\beta$, and
less on $\alpha$.
We do not include any parameterization for mass quenching or mergers in
our simple model (see Peng et al. 2010 for a way in which this may 
be done). For this reason, most of our models overproduce the 
high mass end of the mass function at $z=0$ (which is however subject to some uncertainties, see Bernardi et al. 2010).

The best agreement with the observed evolution of the low-mass end of
the stellar
mass function is found for $\beta \sim -0.1$, i.e. a 
modest negative correlation between sSFR and stellar
  mass. If $\beta \sim$ 0.1, as found in the SAM, the evolution
in the stellar mass function is weak, resembling the weak evolution
found
in the stellar mass function by Guo et al. (2011) and evidenced in our
Figure \ref{fig:evolution}.
Thus \emph{the missing evolution in the stellar mass function
in the SAM is likely due to a positive correlation between 
sSFR and stellar mass}. Models thus need
to find a way to break the close link between specific star formation rates and specific
dark matter accretion rates.

A \emph{negative} correlation between sSFR and stellar
  mass  is a form of
'downsizing' (Cowie et al. 1996). At late times, lower  mass
galaxies are observed to form stars more vigorously relative to their stellar mass
than higher mass galaxies. 
In contrast, 
lower mass haloes are predicted to grow more slowly relative to their mass in a CDM cosmology.  It appears that this form of downsizing, if real, is still not 
explained by current models.

Fig. \ref{fig:toymodel} also shows that $\beta=-0.4$, as found e.g. by Karim et al. (2011), cannot
hold down to low masses, as it would lead to a too rapid evolution
in the stellar mass function. 
A similar conclusion was reached by Drory \& Alvarez (2008), who speculate that  either (i) the excess growth at the low mass end due to
star formation needs to be removed by mergers or (ii) the observational
finding that
$\beta \sim -0.4$ is incorrect and due to 
surveys missing passive low mass galaxies. Also, in agreement with our results, Conroy \& Wechsler (2009)
find that observations of mass growth and star formation at  $z<1$ are roughly
self-consistent, using $\beta \sim -0.2$.

We thus conclude that the problem in the number density evolution in the models is caused by 
a positive, instead of a negative
correlation between sSFR and stellar mass in the models
(see Fig. \ref{fig:tilt}).
We also find that the observed evolution of the low mass end of the mass function is consistent with the 
sSFR-stellar mass relation as observed by Dunne et al. (2009),  Daddi
et al. (2007) and Elbaz et al. (2007), who find $\beta \sim$-0.1.

\subsection{The star formation histories}
\label{sec:sfh}
\begin{figure*}
\centerline{\psfig{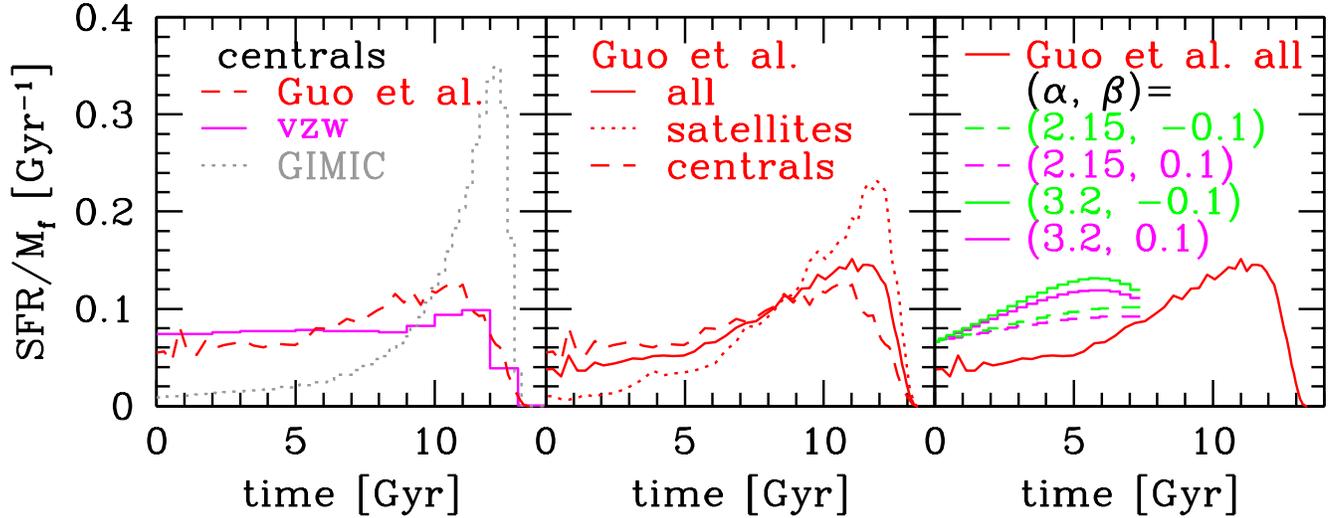}}
\caption{Star formation histories of galaxies from various models in the stellar mass bin 9.27 $<\log(M_{\rm star}/M_{\odot})<$ 9.77, normalized by the total stellar mass ever formed in those galaxies.
Left panel: SFH of central galaxies from the Guo et al. SAM (red dashed line), 
the vzw-simulation (magenta solid line) and the GIMIC simulation (grey dotted line). Middle panel: SFH for all, satellites
and centrals in the SAM shown separately. Right panel:
SFH from toy models with $\alpha$ varying between 2.15 and 3.2, and $\beta$ varying between -0.1 and 0.1, 
for all galaxies. Models 
with $\beta$=0.1 predict insufficient evolution in the MF between $z=1$ and $z=0$, while models with 
$\beta$=-0.1 produce sufficient evolution. Yet, the difference in their SFH is not dramatic.  }
\label{fig:sfh}
\end{figure*}

To establish
 the link between the toy model, and the models discussed in the first part of the paper, 
it is instructive to consider the star formation histories (hereafter SFH)
of galaxies. We define the SFH here as 
the star formation rate of all resolved progenitor galaxies together, divided by the total stellar mass ever
formed. 
In the left panel of Fig. \ref{fig:sfh}, we show the SFH of central
galaxies with 9.27 $<\log(M_{\rm star}/M_{\odot})<$9.77 
in the Guo et al. SAM, GIMIC and in the vzw model.

Reflecting the severe differences between GIMIC on the one hand and the vzw and SAM on the other hand
 seen in Section \ref{sec:approachA},
 the SFH of galaxies in GIMIC is markedly different from the other models. GIMIC predicts a strong initial peak, followed
by a decline, similar as seen in other hydrodynamical models
(e.g. Scannapieco et al. 2012). The vzw simulation, on the other hand, predicts a SFH that
is nearly constant with time in this stellar mass bin, while the SAM
shows a mildly decreasing SFH. 

At first sight, it may seem surprising that such a different SFH as in GIMIC and in the vzw
simulation results in such a similar number
density evolution of galaxies at fixed stellar mass, as seen in Fig. \ref{fig:evolution}.  These results
can however be reconciled if one considers that the two figures refer to different galaxies at $z>0$, and
if one takes into account the evolution of the stellar-to-halo mass ratio, as we explain in sec \ref{sec:recon}.

Of course, to understand the evolution of number densities in Fig. \ref{fig:evolution}, and to compare the GIMIC, vzw model and the SAM
to the toy model, we need to consider the average SFH of all galaxies in those models, and not only the central galaxies. 
 In the middle panel, we
therefore show the SFH of the Guo et al. (2011) SAM galaxies for all,
centrals, and satellites separately. Clearly, satellites have a
different SFH than centrals, lowering the overall SFH at late times.

In the right panel, we again overplot results for all galaxies from
the SAM and compare it to the SFH given by our toy model for
the same stellar mass bin for $\alpha$=2.15 and $\alpha$=3.2, and $\beta$=-0.1 and $\beta$=0.1. Recall
that $\beta$=-0.1 roughly reproduces the observed evolution of the mass function, while $\beta$=0.1 
is very similar to the value in the SAM and leads to too little
evolution. The toy  models do not have a SFH before $z$=0.9 by construction.
Interestingly, the difference between SFH of models that reproduce the number density evolution ($\beta$=-0.1), and those that do not ($\beta$=0.1), is very mild. Only a slight boost in the SFH,
mostly at $z \sim 0.5-1$, is enough to overcome the problem that we have seen
in the number density evolution. The small difference in the SFH might give the impression that the problem in the SAM and the hydrodynamical models
should be easy to solve. This is not necessarily the case. Boosting the sSFR by a small amount
may be difficult, as the necessary gas reservoir has to be in place,
neither having been ejected, stripped (in the case of satellite galaxies) nor having been
converted into stars. We note that
  results for the SFH in our toy model seem to be in reasonable agreement
with the SFH derived by Leitner (2012) using main sequence integration.

\section{Discussion}

\subsection{Two open problems in low mass galaxy evolution}
We find two open problems in low mass galaxy evolution.
First, we confirm a serious discrepancy in the evolution of the number density of galaxies
at fixed stellar mass in semi-analytic and SPH models.

Second, we  show that models do not reproduce the population 
of low mass galaxies with high sSFR and young ages at $z<1$.
The problem becomes more severe towards lower stellar masses. 
Put
  differently, 
as shown in Section \ref{sec:toy}, models
do not reproduce the observed negative slope in the sSFR-stellar
mass relation (see e.g. Somerville et al. 2008). Instead, the slope is positive, which may
  directly cause the too slow evolution in the number density of low
  mass galaxies.
The two problems are thus likely closely connected.

\subsection{Potential explanations by observational issues}
The offset between observations and models at $z>0.5$ could also stem from an incorrect
interpretation of the observations. The two main potential explanations are 
(i) the IMF is variable, leading to problems in the estimates of mass and SFR
and (ii) observations are missing more galaxies than expected. 

Option (i)  is difficult to prove wrong. The observed high sSFR at $z \sim 1-2$ that
are hard to reproduce by models
seem to favour a bottom-light IMF (e.g. Dav\'{e} 2008). If one assumes such a bottom-light IMF, however, this
decreases the observed high redshift mass function even more (see Marchesini et al. 2009), making the discrepancy with models worse. More complex solutions are impossible
  to rule out completely and need to be tested with self-consistent
  models, as a varying IMF will also affect other processes like feedback.

Option (ii) cannot be completely ruled out either, but there are
 several arguments against it. Up to now, no indication of a large and 
unexpectedly faint population of low mass galaxies has been found despite the varying
depth of different surveys. Nevertheless, it is possible that all
surveys have missed a population of very passive and/or dusty low mass galaxies $z>0$.
One argument against this is that such galaxies are not seen in 
large numbers in the models either. For example, in Guo et al. (2011), the fraction of passive galaxies at $z=1$ is only 20 \% in the
mass bin we consider (see Fig. \ref{fig:tilt}, bottom panel.)
Also, observations do not find a large population of passive low mass galaxies at $z$=0 (e.g. Geha et al. 2012).  Even in the local volume, where selection 
effects are minimal, a tight star formation sequence for low mass galaxies has
been found (Lee et al. 2007). It is thus hard to imagine that passive low-mass galaxies would be more numerous at higher redshifts, where the global star formation rate density is higher. Finally, we note that the problem in the number density evolution 
persists up to a mass of $\log(M_{\rm star}/M_{\odot}) \sim $ 10.5 (see e.g. Guo 
et al. 2011), where missing a large number of galaxies becomes more
unlikely. But also if we assume that
observations missing a large number of passive galaxies is the solution to the problem, this would mean that 
star formation does behave very differently than assumed in current models, possibly cycling between 
bursty and passive episodes. This would still constitute a rather fundamental problem for current galaxy formation theory and
would probably require more efficient high $z$ feedback as well (see below).

\subsection{Potential implications}

If the problem with models at $z<1$ is real, which seems likely, a process suppressing galaxy formation at high redshift is needed, so that
the build-up of the mass function below $M^{*}$ can happen at a far later time than
the predicted build-up of the dark matter haloes in which they reside
(see also Conroy \& Wechsler 2009).
Such a process would perhaps also address the problem of the overproduction of satellite
galaxies in models (Weinmann et al. 2006, Lu et al. 2012), as those form early.

Several  mechanisms have been suggested, but all of them seem to have some drawbacks.
Decreasing the star formation efficiency at high redshift leads to an overproduction 
of cold gas in low mass galaxies (e.g. Wang et al. 2012). Preheating the IGM (Mo et al. 2005)
does not seem possible with known mechanisms (Crain et al. 2007). Warm dark matter also does not seem a viable
solution, as we have shown in Section \ref{sec:numden}.  More exotic dark matter candidates like ultra-light dark matter (Marsh et al. 2010) or
mixed dark matter (Boyarski et al. 2009) may need to be considered, but the problem
remains that observations of the Lyman-alpha forest require substantial power on small scales (Seljak et al. 2006).

Thus, perhaps the most natural change to galaxy formation models is changing
the stellar feedback prescription.
In most current models, the efficiency with which gas is
ejected from the cold gas reservoir scales roughly as
\begin{equation}
\dot{M}_{\rm wind}/ \dot{M}_{*}\propto V_{\rm vir}^{-\gamma} \propto  M_{\rm vir}^{-\gamma/3}  t_{\rm H}^{\gamma/3},
\end{equation}
with $M_{\rm vir}$ the halo mass and
  $t_{\rm H}$ the Hubble time. 
Semi-analytic models use $\gamma \sim 2-6$ (Croton et al. 2006, Bower et al. 2006,
Somerville et al. 2008, Guo et al. 2011). 
 In most current hydrodynamical models like GIMIC, 
the initial velocity and mass-loading of the wind is independent of the host halo mass, i.e. $\gamma=0$.
However, Neistein et al. (2012) show that effectively, $\gamma \sim 3/2$ in such simulations, due
to gravitational and hydrodynamical interactions. In the momentum-driven winds (vzw)
model by Oppenheimer
\& Dav\'{e} (2008),  the winds are launched with $\gamma$=1 (with 
the velocity dispersion replacing  $V_{\rm vir}$). The effective $\gamma$ is thus likely 
higher than 3/2. The recently proposed wind scheme by Puchwein \& Springel (2012) that
we included in Fig. \ref{fig:evolution} launches
winds with $\gamma \sim 2$.

As $\gamma>0$ in all these models, 
this  means that their
\emph{star-formation driven feedback at a given halo mass
is less efficient at earlier times}.
As long as  the feedback follows this basic scaling, and as long as star formation 
and cooling become more efficient towards higher redshift, it is 
no surprise that no model is able to predict the steep decrease in the stellar-to-halo
mass ratio towards higher redshifts, which seems demanded by observations (see our Fig. \ref{fig:evolution},
Moster et al. 2010, 2012; Yang et al. 2011). The same problem likely also causes the inability
of the models to reproduce the negative correlation 
between sSFR and stellar mass. In fact, the few
models that we are aware of which reproduce the correlation (the nw model in Dav\'{e} et al. 2011, and 
the no-feedback model in Neistein \& Weinmann 2010) have little or no feedback.
This means that increasing the star-formation driven feedback efficiency arbitrarily will  probably never
solve the basic problem.  What may be needed, therefore, is
a justification to use a different functional form for feedback.

One potential solution might be a more top-heavy IMF at high redshift, which could produce more
massive stars that are short-lived and drive
highly mass-loaded winds. For example, the contribution 
of winds from O and B stars depends strongly on the upper mass cutoff 
of the IMF (Leitherer et al. 1992), but of course, a change in the IMF would 
also affect stellar mass and star formation rate estimates. Alternatively, highly concentrated, clumpy
star-forming regions could also cause higher feedback efficiencies
(Guedes et al. 2011; Brook et al. 2012).

Another potentially important mechanism for understanding the evolution of the 
number density of galaxies is 'reincorporation' or 'GSW recycling' of
material ejected from the galaxy (Oppenheimer et al. 2010). In
semi-analytic models, reincorporation is often assumed to scale either
with  $t_{\rm H}^{-1}$  (Croton et al. 2006), or with $M_{\rm vir}^{-1/3} \times t_{\rm
    H}^{-4/3}$ (Guo et al. 2011). In both cases, it is thus most important at early
  times. This means that the net efficiency of outflow of mass from a given host halo, 
which is the outflow efficiency minus the reincorporation efficiency,
is low at high
redshift in these models for two reasons: First, the feedback efficiency is low, and second, 
the reincorporation efficiency is high.
At late times, when it could boost star formation rates, it is least
efficient.
This basic scaling used in semi-analytic models thus works against solving the problem. Also, 
it is not confirmed by SPH simulations: Oppenheimer et al. (2010) have found
a much weaker dependence on time, which makes recycling relatively
more important at late times, and probably explains why the vzw
model produces an almost constant SFH at $\log(M_{\rm star}/M_{\odot})\sim 9.5$. On the other hand,  GSW recycling
is  more efficient for high mass haloes than for low mass haloes, which exacerbates
the problem of the incorrect relation between sSFR and stellar mass (Firmani, Avila-Reese \&
Rodr\'{i}guez-Puebla 2010).

\subsection{Reconciling the star formation histories and number density evolution}
\begin{figure*}
\centerline{\psfig{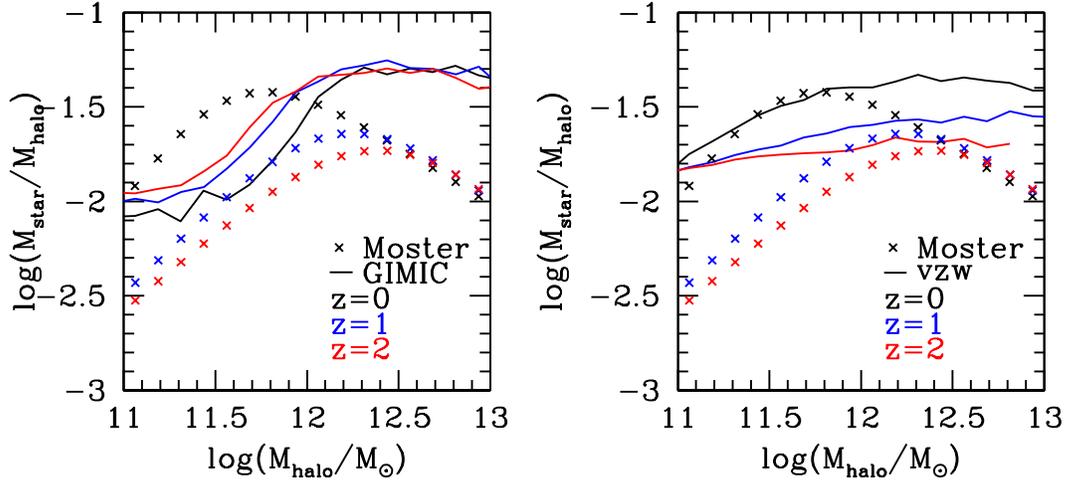}}
\caption{The stellar-to-halo mass ratio as a function of halo mass. Crosses show abundance matching results from 
Moster et al. (2012).
On the left, results from GIMIC are overplotted, on the right, from vzw.  $z=0$ results are shown in black, $z=1$ results in blue and $z=2$ results in red. }
\label{fig:msmh}
\end{figure*}

\label{sec:recon}
One interesting result of our study is that the number density evolution of low mass galaxies is very 
similar in all models (Fig. \ref{fig:evolution}), although the star formation rates, ages (Fig. \ref{fig:sfr}) and
star formation histories (Fig. \ref{fig:sfh}) of the galaxies which populate the mass bin at $z=0$
are very different.

To link these results, it is instructive to consider the stellar-to-halo mass ratio of galaxies
as a function of halo mass. 
We show this quantity as calculated by Moster et al. (2012) from abundance matching, compared
to central galaxies in GIMIC and vzw, in Fig. \ref{fig:msmh},  with colours denoting $z=0$ (black), 
$z=1$ (blue) and $z=2$ (red).

Two interesting points can be noted from this figure. 
First, the stellar-to-halo mass ratio of haloes with mass $\log(M_{\rm halo}/M_{\odot}) \sim 11.25$, which host
the galaxies in our stellar mass bin at $z=0$, evolves very little up to $z=2$ both in the 
vzw and GIMIC simulations. This explains the
fact that the number density evolution is fairly similar in those models in Fig. \ref{fig:evolution}. Observations, on the other hand, seem to demand a much 
stronger evolution in the stellar-to-halo mass ratio since $z=1$ at these halo masses.

Fig. \ref{fig:msmh} also helps us to understand the large differences in the SFH 
between GIMIC, vzw and what we have derived from our toy model in Fig. \ref{fig:sfh}. 
Assume that in all cases we start with a halo of mass $\log(M_{\rm
  halo}/M_{\odot})$ = 11.1 at $z=1$, growing to 11.3 at $z=0$.
In GIMIC, the mass of the central galaxy will only grow by 20 \%, since 
the stellar-to-halo mass
ratio actually decreases from $z=1$ to $z=0$, thus requiring only very little star formation.
 In vzw, on the other hand, the central galaxy needs to grow by a factor of 2.5 in the same period, 
in good agreement with the much higher late SFR in this model.
Looking at the Moster et al. (2012) results, the same galaxy needs to grow by a factor of about 8, 
obviously requiring even more late star formation, well
in line with the results from our toy model in Fig. \ref{fig:sfh}.

\section{Conclusions}

We have investigated the dramatic failure of models to predict the number density evolution of
galaxies with stellar masses of about $\log(M_{\rm star}/M_{\odot})$
$\sim$ 9.5. For this, we have used two approaches. First, we have compared the $z=0$ properties of
low mass galaxies in models and observations, using two different semi-analytic models and
two different hydrodynamical simulations, and excluding satellite galaxies. 
 Second, we have built a simple toy model to investigate the link between
the sSFR as a function of redshift and stellar mass, and the evolution of the mass function.

Our main findings are as follows.
\begin{itemize}
\item We confirm the potentially serious problem in the number density evolution 
of galaxies with $\log(M_{\rm star}/M_{\odot})$
$\sim$ 9.5. The observed evolution 
in the number of galaxies at fixed stellar mass 
is much steeper than in the models. This indicates that the
growth of galaxies at high redshift is too efficient in the models,
causing galaxies to be in place too early.  We show
that this problem does not only appear in SAMs, but also, in 
remarkably similar form, in SPH simulations, indicating that it is a real 
problem in the current theory of galaxy evolution. We also show
that the number density evolution of model galaxies closely matches
that of dark matter haloes hosting these galaxies at $z=0$. Such a close
correspondence between galaxies and dark matter haloes seems to be
missing in the real Universe, if we assume a $\Lambda$CDM cosmology.
We have also found that assuming a WDM instead of CDM cosmology gives a very similar
number density evolution of dark matter haloes, which indicates that WDM will
likely not help to solve the problem we describe in this work.
\item This problem in the number density evolution
is likely related to the fact that the models fail to reproduce
the 40 \%          of central galaxies with       luminosity-weighted ages below 2 Gyr, and the 10 \% 
with 
star formation rates in excess of log(sSFR)=-9.5 at $z=0$ in our
  stellar mass bin. We find that a similar problem appears at $z=1$, and
that the problem seems to become more severe towards lower stellar masses.
\emph{It seems that many low mass galaxies have experienced substantial recent growth, which 
is a phenomenon  not seen in models.}
\item In order to understand the potential link between the evolution of
number densities and the sSFR-stellar mass relation, we build a simple toy model.
We find that the evolution of the stellar mass function with 
time is very sensitive to the assumed slope in the sSFR-stellar
  mass relation, $\beta$,
with sSFR $\propto M_{\rm star}^{\beta}$. In models, $\beta$ 
is usually positive, causing a very slow evolution of the low-mass end of the stellar mass function. The relatively fast observed evolution of the
 stellar mass function, on the other hand, favours a negative beta,
 with $\beta \sim
-0.1$. This means that the observations advocating $\beta \sim -0.1$ seem consistent
with the observed evolution of the mass function. We point out that $\beta \sim
-0.4$, as found by some observational studies directly measuring the
sSFR, would cause a too strong evolution in the mass function,
confirming results by Drory \& Alvarez (2008).
\item The inability of the models to reproduce the number
density evolution of galaxies, the population of young and star forming galaxies and the
negative correlation between sSFR and stellar mass seem all to be part of the 
same underlying problem: Despite the presence of strong stellar feedback, model galaxies closely follow the evolution of
dark matter haloes in a $\Lambda$CDM cosmology. As low mass dark matter haloes form earliest
and evolve the least in a hierarchical cosmology, low mass galaxies also come out
old and evolved today, in clear contrast with observational results. 
It is thus necessary to find a way to decouple the halo accretion rate
and the star formation 
rate of low mass galaxies.

Finally, the failure of all models to suppress
galaxy formation at high redshift is likely
caused by the fact  
that most feedback prescriptions are essentially of the same flavor in these models. The feedback efficiency in most current models
scales as $M_{\rm vir}^{-\gamma/3} t_H^{\gamma/3}$ where $\gamma \geq 0$, resulting in a reduced feedback efficiency 
for a given halo mass at earlier time. The overproduction of low-mass galaxies at $z>0.5$ in all SAMs and hydrodynamical simulations explored here may
thus be symptomatic of the limited range of feedback prescriptions
currently in use, and indicates that alternative models of feedback
need to be explored.  Other potential remedies include metallicity-dependent
star formation laws, an evolving IMF,
observations missing an unexpectedly
 large number of galaxies at high redshift, or perhaps some form of pre-heating.

\end{itemize}

\section*{Acknowledgments}
We acknowledge funding from ERC grant HIGHZ no. 227749. 
We thank Pablo P\'{e}rez-Gonz\'{a}lez, David Gilbank, Jarle Brinchmann, Dave Wilman, Sean McGee, Rita Tojeiro,
Anna Gallazzi, Britt Lundgren, Mattia Fumagalli and Gabriel Brammer
for their kind assistance 
with their observational data (not all of which ended being used in the final 
version of the paper). We thank Jie Wang for making his
semi-analytic model available and Gabriella De Lucia for computing
luminosity-weighted ages in this model, and Ewald Puchwein for
providing the data for their simulation. We thank
the referee for suggestions which helped to improve the paper, 
and Joop Schaye for detailed and helpful comments on the manuscript.
We also thank Romeel Dav\'{e}, Sandy Faber, Gabriella De Lucia, Simon
Lilly, Ryan Quadri, Marijn Franx, Joop Schaye and
Christian Thalmann for useful discussion.

     SQL databases containing the full galaxy data for the SAM of G11
    at all redshifts and for both the Millennium and Millennium-II 
    simulations are publicly released at 
    \texttt{http://www.mpa-garching.mpg.de/millennium}. The Millennium site was created as 
    part of the activities of the German
    Astrophysical Virtual Observatory. 

The GIMIC simulations were carried out by the Virgo Consortium using the HPCx facility at the Edinburgh Parallel Computing Centre, the Cosmology Machine at the University of Durham, and on the Darwin facility at the University of Cambridge.

    Funding for the SDSS and SDSS-II has been provided by the Alfred
    P.\ Sloan Foundation, the Participating Institutions, the National
    Science Foundation, the U.S.\ Department of Energy, the National
    Aeronautics and Space Administration, the Japanese Monbukagakusho,
    the Max Planck Society, and the Higher Education Funding Council
    for England. The SDSS Web Site is \texttt{http://www.sdss.org/}. 
    The SDSS is managed by the Astrophysical Research Consortium for
    the Participating Institutions. The Participating Institutions are
    the American Museum of Natural History, Astrophysical Institute
    Potsdam, University of Basel, University of Cambridge, Case
    Western Reserve University, University of Chicago, Drexel
    University, Fermilab, the Institute for Advanced Study, the Japan
    Participation Group, Johns Hopkins University, the Joint Institute
    for Nuclear Astrophysics, the Kavli Institute for Particle
    Astrophysics and Cosmology, the Korean Scientist Group, the
    Chinese Academy of Sciences (LAMOST), Los Alamos National
    Laboratory, the Max-Planck-Institute for Astronomy (MPIA), the
    Max-Planck-Institute for Astrophysics (MPA), New Mexico State
    University, Ohio State University, University of Pittsburgh,
    University of Portsmouth, Princeton University, the United States
    Naval Observatory, and the University of Washington.

\appendix

\section{Additional comparisons}
Below, we compare the dust attenuations and stellar metallicities
between models and observations in
our stellar mass bin at $z=0$. We find that agreement is
reasonable, except that SAMs underpredict the dust attenuation in galaxies.

\subsection{Stellar metallicities}

\begin{figure*}
\centerline{\psfig{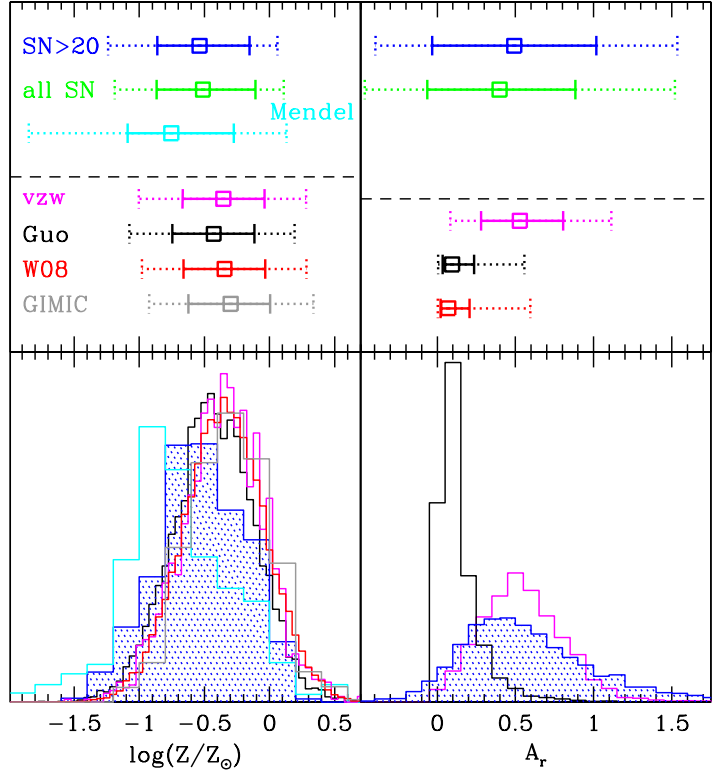}}
\caption{Model and observational results for the stellar metallicities
  (left  panels) and dust attenuations (right  panels) in the stellar mass bin $\log(M_{\rm star}/M_{\odot})=9.27-9.77$.  Top panels
show the median value (empty square), the range within which 68 \% of the values lie (solid errorbars) and
the range within which 95 \% of the values lie (dotted errorbars). Bottom panels show the full distributions
for some selected datasets. The colour coding in the left  panel is as follows.
 Blue and green: Gallazzi et al. (2005) luminosity-weighted
metallicities for the $S/N>20$ and the full sample. Cyan: Mendel et al. sample. Magenta: vzw simulation. Red: Wang et al. (2008)
SAM. Grey: GIMIC.
Blue and green: dust attenuation following Kauffmann et al. (2003), S/N$>$20 and the full sample. 
Magenta: vzw simulation. Black: Guo et al. SAM. Red: Wang et al. (2008). }
\label{fig:age}
\end{figure*}

In Fig. \ref{fig:age}, left  panels, we show the distribution of stellar metallicities
for our various samples. The metallicities are luminosity-weighted for the observed sample [blue
and cyan lines for the Gallazzi et al. (2005) and Mendel et al. estimates, respectively] but 
stellar mass-weighted for all the models. For the SAMs and GIMIC, total metallicities are a direct model output, 
while they are calculated from the Fe and O abundance for the vzw model, and computed
from spectral indices in the observational data.

We convolve all model results with a Gaussian distribution with $\sigma$=0.3 in log($Z$)
which is the mean 68 \% confidence range as indicated in Gallazzi et al. (2005) for
 the corresponding stellar mass bin. 
Interestingly, metallicities from the Wang et al. (2008) model, 
the Guo et al. (2011) model and the vzw simulation are in near perfect agreement with 
each other, indicating
that the predictions of models are robust in this respect. They are also 
in reasonable agreement with observations.
The remaining
relatively small offset between models and observations could be due to the difference between mass-weighted
and luminosity-weighted metallicities.
Dav\'{e} et al. (2011b) obtain qualitatively similar results when comparing gas-phase metallicities
in the vzw simulation to observational results.

\subsection{Dust}
In Fig. \ref{fig:age}, right panel, we show the distribution of $r$-band dust attenuation in models and
observations.
Both semi-analytic models
seem to include too little dust attenuation. On the other hand, the vzw model predicts dust attenuations in
good agreement with the SDSS estimates. This is not surprising as dust attenuation in this model
is included following the observed relation between metallicity and
dust attenuation. 
It is interesting that the Guo et al. (2011) SAM manages to reproduce the observed stellar mass function and
the $r$-band luminosity function very well (see Guo et al. 2011) despite underestimating dust attenuation. This
might be due to the fact that the SAM \emph{also} seems to produce too old, and perhaps slightly too metal-rich 
galaxies. This will tend to make galaxies at fixed stellar mass less luminous and redder, thus partially compensating for
the under-estimate in the dust attenuation. 
The large discrepancy in the dust attenuation between the SAMs and the observations is a warning that one
should prefer using direct physical quantities like sSFR over colour when doing model-observation
comparisons.

\end{document}